# Combined computational and experimental investigation of the $La_2CuO_{4-x}S_x$ ($0 \le x \le 4$) quaternary system

Hua He,[1]* Chuck-Hou Yee,[2] Daniel E. McNally,[3] Jack W. Simonson,[4] Shelby Zellman,[1] Mason Klemm,[1] Plamen Kamenov,[3] Gayle Geschwind,[3] Ashley Zebro,[3] Sanjit Ghose,[5] Jianming Bai,[5] Eric Dooryhee,[5] Gabriel Kotliar,[2] Meigan C. Aronson[1]

[1] Department of Physics and Astronomy, Texas A&M University, College Station, TX 77843, USA

[2] Department of Physics and Astronomy, Rutgers University, Piscataway, NJ 08854, USA

[3] Department of Physics and Astronomy, Stony Brook University, Stony Brook, NY 11794, USA

[4] Department of Physics, Farmingdale State College, Farmingdale, NY 11735, USA

[5] National Synchrotron Light Source II, Brookhaven National Laboratory, Upton, NY 11973, USA

* maronson@tamu.edu

Abstract: The lack of a mechanistic framework for chemical reactions forming inorganic extended solids presents a challenge to accelerated materials discovery. We demonstrate here a combined computational and experimental methodology to tackle this problem, in which in situ X-ray diffraction measurements monitor solid state reactions and deduce reaction pathways, while theoretical computations rationalize reaction energetics. The method has been applied to the $La_2CuO_{4-x}S_x$ ($0 \le x \le 4$) quaternary system, following an earlier prediction that enhanced superconductivity could be found in these of new lanthanum copper(II) oxysulfide compounds. In situ diffraction measurements show that reactants containing Cu(II) and S(2-) ions undergo redox reactions, leaving their ions in oxidation states that are incompatible with forming the desired new compounds. Computations of the reaction energies confirm that the observed synthetic pathways are indeed favored over those that would hypothetically form the suggested compounds. The consistency between computation and experiment in the $La_2CuO_{4-x}S_x$ system suggests a new role for predictive theory: to identify and to explicate new synthetic routes for forming predicted compounds.

**Statement of Significance**

Discovery of new materials enabling new technologies, from novel electronics to better magnets, has so far relied on serendipity. Computational advances show promise that new materials can be designed in a computer and not in the lab, a proposal called `Materials by Design'. We present here a detailed comparison between theory and experiment, carrying out the synthesis of a high temperature superconductor in an x-ray beam to elucidate the sequence of chemical reactions as the compound forms. Parallel computations of the stabilities of possible compounds that could form from the selected elements accurately predict the observed reactions. Paired with our chemical intuition, this methodology provides understanding and potentially control of the essential chemical principles responsible for stabilizing virtually any compound.

The discovery of new solid inorganic materials is the primary driver for advancing our understanding of the formation of extended solids and their functional properties. While chemical guidelines like isoelectronic substitution and metathesis reaction have had some success in guiding exploratory syntheses[1-3], the discovery of new solid inorganic materials so far remains largely serendipitous. Advancing beyond chemical intuition has proven difficult. Unlike organic syntheses, in which only specific sites of a molecule are modified in chemical reactions, the synthesis of extended solids involves the breakdown and reassembly of entire atomic lattices, and there is no mechanistic framework to describe these processes yet[4]. The development of a set of general rules that clarify the essential chemical reactions governing the assembly of atoms in extended solids remains a central goal of materials-inspired research. First-principles calculations such as density functional theory (DFT) calculations[5,6] are becoming increasing accurate in computing the energies of different structures and their heats of formation, crucial steps towards realizing the challenging goal of designing new materials with desired functionality without prior knowledge of the crystal structure[7-13]. However, these predicted compounds have too often eluded experimental discovery, even when they are predicted to be thermodynamically stable. We envisage here a new way that predictive theory can accelerate the discovery of new compounds, by clarifying the energetics of the different steps of the reactions that comprise candidate syntheses, and ultimately using this insight to recommend syntheses that are likely to be viable. Previously, there was a decided lack of direct information about reactions leading to the formation of extended solids that could subsequently be compared to theory. The use of in situ X-ray diffraction (XRD) experiments, where the syntheses are carried out in a high energy X-ray beam, provides a powerful way to explore these chemical reactions[14-16]. The formation of a crystalline phase can be observed in real time, providing a record of the sequence of structures and phases that occur on heating and cooling. We will show here that by using this information, one can deduce the reaction pathways and rationalize the energetics of the chemical reactions, establishing a step-by-step correspondence between computations and experiments. The goal is to connect our chemical intuition to predictive theory, and in this way to gain insight into reaction pathways that are central to the formation of extended solids. An added benefit is that the exploration of materials phase diagrams, a foundational task for materials-inspired research, is greatly accelerated using in situ X-ray diffraction measurements.

Here we present our investigation of the $La_2CuO_{4-x}S_x$ ($0 \leq x \leq 4$) quaternary system using experimental and computational methods in parallel to show the advantages of this combined method. Previously we have proposed theoretical compounds $La_2CuO_3S$ and $La_2CuO_2S_2$[17], which are obtained by replacing half or all of the apical oxygen with sulfur in $La_2CuO_4$ (Fig. 1), the parent compound of the first high-temperature superconductor discovered within the cuprate family[18]. Electronic structure calculations indicate that sulfur substitution produces large effects on the charge-transfer energy and orbital distillation, providing a concrete mechanism to distinguish between competing theories of high-Tc superconductivity[17,19-21]. Nevertheless, experimental realization of $La_2CuO_2S_2$ or $La_2CuO_3S$ has never been reported, nor any related oxysulfide containing copper(II). Here we use in situ XRD measurements to examine this $La_2CuO_{4-x}S_x$ quaternary system to determine whether $La_2CuO_3S$ and $La_2CuO_2S_2$, which were predicted to be locally stable but globally thermodynamically unstable[17], actually form, and if so, by which synthetic route. Our approach is to explicate the range of compounds that are possible as sulfur is introduced into $La_2CuO_4$, to determine whether S replaces O as a dopant, and at larger levels, whether $La_2CuO_3S$ and $La_2CuO_2S_2$, or perhaps other compounds actually form experimentally. Establishing a quaternary phase diagram connecting $La_2CuO_4$ and $La_2CuS_4$ is a substantial undertaking, and we will demonstrate here how effective in situ XRD measurements are for accelerating this foundational synthetic task. An equally important aspect of the in situ approach is that it allows us to monitor the progress of the chemical reactions in detail, and to deduce reaction pathways under the actual experimental conditions. Combined with theory, this information opens the door to a new role for predictive theory in exploratory synthesis. The increasing availability of databases containing computational results on a wide range of materials, including their heats of formation, allow us the opportunity to compare the stabilities of the proposed target compositions against those of known compounds in the La-Cu-O-S chemical system. In this way, we aim to understand the energetics of the experimentally-observed reactions. Combining the theoretical and experimental studies, we have been able to explicate the experimental challenges incumbent in synthesizing the theoretical compounds $La_2CuO_3S$ and $La_2CuO_2S_2$, or in general, copper(II) oxysulfides.

## Results and discussion

We first conducted the synthesis of the parent compound $La_2CuO_4$ following the conventional solid state reaction method[22]. The mixture of $La_2O_3$ and CuO powders was heated up to high temperature while the reaction was monitored using in situ XRD as shown in Fig. 2. The diffraction patterns at room temperature suggest that instead of $La_2O_3$, $La(OH)_3$ is present as the starting material, which is not surprising since $La_2O_3$ is hygroscopic and it readily absorbs moisture from the uncured high temperature cement, forming $La(OH)_3$. Nevertheless, the presence of $La(OH)_3$ doesn’t seem to affect the formation of $La_2CuO_4$. At elevated temperature, $La(OH)_3$ gradually loses water, first forming LaOOH, and then turning back to dry $La_2O_3$, as shown by the diffraction patterns. The formation of $La_2CuO_4$ is observed at temperatures above 860°C, though it grows very slowly at this temperature. Raising the temperature above 1000°C greatly accelerates the growth of $La_2CuO_4$, which is most likely due to the much faster diffusion of CuO at these temperatures[23]. These observations are consistent with literature reports in which polycrystalline $La_2CuO_4$ powder is usually synthesized by firing $La_2O_3$ and CuO powders at temperature between 900°C and 1100°C[24,25].

The energy of formation of $La_2CuO_4$ from $La_2O_3$ and CuO ($\Delta E = E_{\text{products}} - E_{reactants}$) has been evaluated from the difference of their total energies (Table 1). We note that although mixing $La_2O_3$ and CuO at elevated temperature is a well-established method for synthesizing $La_2CuO_4$, the computed reaction energy is actually positive[17], implying that the product is unstable with respect to the binary oxides. It is plausible that the formation of $La_2CuO_4$ is driven by kinetics rather than the thermodynamic driving force, i.e., the growth of $La_2CuO_4$ crystallites is faster than its decomposition at elevated temperature. On the other hand, the small positive reaction energy, 51 meV/atom, could simply represent the inherent uncertainty of this type of computational methodology. Indeed, estimates of the standard deviation of the difference between computed and experimental reaction energies are on the order of 50 meV/atom[26,27]. Given the accuracy of these DFT calculations, it seems likely that $La_2CuO_4$ is on the verge of being thermodynamically stable.

Following the procedures established for synthesizing the parent compound $La_2CuO_4$, we employ the same method to examine the solid state reactions toward the theoretical compound $La_2CuO_3S$. We first investigate the reaction pathway $La_2O_2S$+CuO=$La_2CuO_3S$, noted as Proposal A in Table 1. As shown in Fig. 3a the diffraction patterns remained unchanged up to ~760°C and above that temperature, new diffraction peaks gradually appeared, e.g. peaks at Q~1.6 $Å^{-1}$ (double peaks) and 2.1 $Å^{-1}$ corresponding to $La_2O_2SO_4$ and the peak at Q~2.58 $Å^{-1}$ corresponding to $Cu_2O$, concomitant with the weakening of the CuO and $La_2O_2S$ peaks. The weight fractions obtained from Rietveld refinements also indicate the gradual increase of $La_2O_2SO_4$ and $Cu_2O$, and the decrease of $La_2O_2S$ and CuO over the same temperature range. Above 1000°C, additional chemical reactions occurred, as new diffraction peaks that could be identified as belonging to LaCuSO and $La_2O_3$ appeared, while the starting material $La_2O_2S$ and the intermediate phase $Cu_2O$ were almost depleted. Based on the evolution of the phases, we are able to draw a step-wise reaction pathway for the reactions between $La_2O_2S$ and CuO at high temperature. As shown in Table 1, $La_2O_2S$ and CuO first react at ~760°C, forming $La_2O_2SO_4$ and $Cu_2O$. At higher temperature, the newly formed $Cu_2O$ reacts with the remaining $La_2O_2S$, forming LaCuSO and $La_2O_3$. The computed energies of these two experimental reactions are both negative, indicating these are thermodynamically favorable reactions. Reaction step 1 where $La_2O_2S$ + 8 CuO = 4 $Cu_2O$ + $La_2O_2SO_4$ has an especially negative energy –197 meV/atom, which provides a large thermodynamic driving force for the forward reaction, compared to the proposed reaction $La_2O_2S$ + CuO = $La_2CuO_3S$, which has instead a large positive energy 182 meV/atom.

It is important to point out that the chemical equations reported here may not represent the entirety of the reactions that actually occur, since XRD is only sensitive to crystalline materials. Reactions producing amorphous products or gases, or reactions involving moisture and/or oxygen that were accidentally introduced into the sample container may not be entirely captured, and may sometimes need to be deduced. For example, the experiment actually produced slightly more $La_2O_2SO_4$ and less LaCuSO compared to the amounts that could have been made if the equations were strictly followed. This discrepancy could be readily explained if a small amount of $La_2O_2S$ were oxidized to $La_2O_2SO_4$ by $O_2$ in the sample tube, instead of by CuO (Supplementary Fig. 2). Nevertheless, the evolution of the diffraction patterns provides a clear and understandable account of the main reactions in the $La_2CuO_{4-x}S_x$ system.

We then tested the second theoretical reaction, Proposal B, where $La_2O_3$ and CuS were combined to form the theoretical compound $La_2CuO_3S$. As seen from Fig. 3b at temperatures near 450°C, we notice the decomposition of CuS into $Cu_2S$, followed by the appearance of $La_2O_2S$ and $La_2O_2SO_4$. We propose that S released from the decomposition of CuS reacts immediately with $La_2O_3$, forming $La_2O_2S$ and $La_2O_2SO_4$. This proposal has been subsequently confirmed by conventional lab-based synthesis where S and $La_2O_3$ powders were combined, showing that the same products formed at high temperatures. At temperatures above 860°C, the intermediate phases $La_2O_2S$ and $Cu_2S$ react, forming LaCuSO. The step-wise reaction pathways and the total reaction are summarized in Table 1 together with their computed reaction energies. Both experimental steps, 8 CuS + 4 $La_2O_3$ = 4 $Cu_2S$ + 3 $La_2O_2S$ + $La_2O_2SO_4$ and $La_2O_2S$ + $Cu_2S$ = 2 LaCuSO, have negative reaction energies which provide the driving force. Overall, the total reaction energy of this two-step process is negative, in contrast with the positive 266 meV/atom for the direct reaction $La_2O_3$ + CuS = $La_2CuO_3S$ that was theoretically proposed.

Synthesis of the suggested compound $La_2CuO_3S$ was also attempted using $La_2S_3$ as a sulfur source (Proposal C). From Fig. 3c, it is clear that at temperatures between 600 and 760°C, the weight fractions of $La_2S_3$, $La_2O_3$ and CuO drop simultaneously, concomitant with the appearance of the additional phases $Cu_2O$, $La_2O_2S$ and $La_2O_2SO_4$. It is not clear whether $La_2S_3$ is oxidized directly by CuO, or if $La_2S_3$ first reacts with $La_2O_3$ to form $La_2O_2S$, and then $La_2O_2S$ is subsequently oxidized by CuO, since the additional phases appear simultaneously. Nevertheless, the net reaction is the same for both reaction pathways, and they both have large negative reaction energies, as shown in Table 1. Above 760°C, LaCuSO is formed by the reaction between $La_2O_2S$ and $Cu_2O$, the same reaction as found in experiment A. Again, the total reaction has large negative reaction energy, –221 meV/atom, in sharp contrast with the positive 79 meV/atom involved with the theoretical proposed reaction.

Our experiments so far indicate that the theoretically proposed syntheses for the formation of $La_2CuO_3S$ are not realistic, and we have been able to use total energy calculations to explain the energetically favorable outcomes that are observed in our experiments. Consequently, we adopted a broader charge for our synthesis program, which is the examination of the compositional space, $La_2CuO_{4-x}S_x$ ($0 \leq x \leq 4$), with several aims: (1) if smaller amounts of S could be doped into $La_2CuO_4$; (2) if higher concentration of S in starting materials would facilitate the formation of the theoretically proposed compounds; (3) the existence of any previously unreported ternary or quaternary compound. Eight different compositions ($x$=0, 0.25, 0.5, 0.75, 1, 2, 3, and 4) have been screened using the in situ XRD method, and the products formed at different temperatures are identified and then used in the construction of the non-equilibrium phase diagram displayed in Fig. 4. In general, no chemical reaction occurs at temperatures below ~600°C (the light purple region), except the changes among $La_2O_3$, $La(OH)_3$, and LaOOH that we described earlier. As the temperature is raised above 600°C, the reactions between $La_2S_3$ and CuO start, producing $Cu_2O$ and $La_2O_2SO_4$ as indicated in the yellow region in Fig. 4. The final products produced at higher temperatures depend on the sulfur concentrations $x$. In the O-rich region (green region, $x$<0.25), the $La_2CuO_4$ phase is still among the list of products. However, refinements of the $x$=0.25 data suggest that essentially all S has ended up in the product $La_2O_2SO_4$, leaving none available for S-doping in $La_2CuO_4$. The quaternary product LaCuSO prevails in the region of 0.75<$x$<3 (blue region), and when $x$ is close to 3, additional phases $LaCuS_2$[28] and $La_3CuO_2S_3$[29] appear, and when $x$=4, $LaCuS_2$ and $La_2CuS_4$[30] phases dominate. Except for these overall observations, side-by-side comparisons of the reactions have also revealed several interesting aspects of this system. First, the oxidized product $La_2O_2SO_4$ doesn't survive for the full S range. When the sulfur concentration $x$<2, $La_2O_2SO_4$ grows into a better crystalline product at temperature above 800°C (sharper peaks, as indicated by the peak split at Q~1.6 Å), and when $x \geq 2$, $La_2O_2SO_4$ transforms into $La_2O_2S$ above 800°C. We speculate that the formation of $La_2O_2SO_4$ involves two steps, where in the first step only an intermediate phase forms (noted as $La_2O_2SO_4$* in Fig. 4) and the bona fide $La_2O_2SO_4$ forms in the second step during which the participation of $La_2O_3$ is required, as indicated by the chemical equations in Experiment C. We conclude that the availability of $La_2O_3$ at above 800°C determines whether the final product is $La_2O_2SO_4$ or $La_2O_2S$. The product lists shown in Supplementary Table 2 clearly support this argument. Second, a different reaction pathway between $La_2O_2S$ and $Cu_2O$ has been revealed by the experiment with $x$=0.5. Instead of forming LaCuSO, we have observed the formation of Cu and $La_2O_2SO_4$ from the reaction between $La_2O_2S$ and $Cu_2O$ in this experiment. The two different chemical equations and their reaction energies are determined to be

$$2\,La_2O_2S + Cu_2O = 2\,LaCuSO + La_2O_3 \ (\Delta E = -54 \text{ meV/atom}) \quad (1)$$

$$La_2O_2S + 4\,Cu_2O = 8\,Cu + La_2O_2SO_4 \ (\Delta E = -223 \text{ meV/atom}) \quad (2)$$

The validation of both reactions has been further corroborated by conventional lab syntheses (Supplementary Fig. 3). The relative magnitudes of the reaction energies suggest that the second reaction should always dominate. However, considering a starting composition of $La_2O_2S$ and $Cu_2O$ in a 2:1 molar ratio, a large amount of $La_2O_2S$ would remain unreacted if the reaction followed equation (2). This will actually alter the reaction energy as follows:

$$2\ La_2O_2S + Cu_2O = 2\ Cu + \frac{1}{4}\ La_2O_2SO_4 + \frac{7}{4}\ La_2O_2S\ (\Delta E = -73\ \text{meV/atom})$$

Thus, we see that this reaction energy then is the same as that of equation (1), taking into account of the accuracy of the DFT computations. This explains the formation of LaCuSO when the molar ratio of the starting compositions $La_2O_2S/Cu_2O$ is close or larger than 2, which is the case for $x$>0.75.

Although multiple synthetic routes have been tested, it has not been possible to synthesize the theoretical compound $La_2CuO_3S$ or $La_2CuO_2S_2$, seemingly due to the formation of competing phases, namely LaCuSO and $La_2O_2SO_4$. To thoroughly evaluate the thermodynamic stabilities of the theoretical compounds and the competing phases, we have obtained the total energies of the related elements, binary, ternary and quaternary compounds. From this information we have constructed the convex hull, which is the locus of formation energies for both known and theoretical compounds. The thermodynamic stability of a compound can be evaluated by its energy relative to the hull, i.e., the energy difference between a compound and the set of the most stable compounds with the same averaged chemical composition[31]. A positive energy above the hull indicates that the compound is unstable, and there is a thermodynamic driving force for it to decompose into two or more stable compounds. Conversely, a negative energy below the hull suggests a stable compound. The blue points in Fig. 5a present the known compounds in the La-Cu-O-S quaternary system, which all have negative energies. The theoretical compounds $La_2CuO_3S$ and $La_2CuO_2S_2$ have large positive energies above the hull, 324 meV/atom (by decomposing into $La_2O_2S$, Cu, and $La_2O_2SO_4$) and 232 meV/atom (by decomposing into $La_2O_2S$ and CuS), suggesting that both are thermodynamically unstable. As expected, the products that appear in our syntheses are stable, having a negative energy with respect to the hull (Fig. 5b). While metastable solid phases such as $La_2CuO_4$[17] and $La_8Cu_7O_{19}$[32] certainly exist and can be readily synthesized, these compounds are in general computed to have total energies that are about 50 meV/atom above the hull energy, in sharp contrast with the large values for $La_2CuO_3S$ and $La_2CuO_2S_2$.

The ability to track the reaction pathways that culminate in the formation of $La_2CuO_4$ and LaCuSO at high temperatures, and the decisive absence of the theoretically proposed compounds $La_2CuO_3S$ and $La_2CuO_2S_2$, make it very clear that the underlying mechanism that controls the reactions is a redox reaction that transforms $Cu^{2+}$ to $Cu^{1+}$, or even $Cu^0$, in the presence of $S^{2-}$. While the experimental reaction pathways shown in Table 1 all have a reaction energy that is negative, the initial, low-temperature steps of the chemical reactions generally have particularly large negative reaction energies (–197 meV/atom in Experiment A and –231 meV/atom in Experiment C), suggesting a large thermodynamic driving force for these reactions. By examining these reactions we notice that these are redox reactions between $Cu^{2+}$ and $S^{2-}$ containing compounds, i.e. $La_2O_2S + 8\ CuO = 4\ Cu_2O + La_2O_2SO_4$ for experiment A and $8\ CuS + 4\ La_2O_3 = 4\ Cu_2S + 3\ La_2O_2S + La_2O_2SO_4$ for experiment B. At intermediate temperatures (500-760°C), well below the formation temperature of $La_2CuO_4$, the $Cu^{2+}$-containing compound reacts with the $S^{2-}$-containing compound, changing their oxidation states to $Cu^{1+}$ and $S^{6+}$ (as in sulfate $SO_4^{2-}$), respectively. Note that for the proposed $La_2CuO_3S$ and $La_2CuO_2S_2$, the oxidation state of Cu is $Cu^{2+}$ and the oxidation state of sulfur is $S^{2-}$. The inability to realize these desired oxidation states seems to be the essential reason why S-doped $La_2CuO_4$ cannot be made by high temperature solid state synthesis. This suggests that the most successful approaches to synthesize compounds like $La_2CuO_3S$ will take place at low temperatures where the redox reaction between $Cu^{2+}$ and $S^{2-}$ can be avoided. In that sense, low-temperature synthesis, for example, introducing a flux or hydrothermal synthesis may be a more appropriate synthetic route[33,34].

The overwhelming weight of experimental and theoretical evidence suggests that the theoretical compounds $La_2CuO_3S$ and $La_2CuO_2S_2$ are unlikely to be realized via conventional solid state synthesis[22], due to their formidable thermodynamic instabilities. At the same time, our experiments demonstrate most eloquently the power of in situ XRD synthesis as a way to articulate reaction mechanisms on a step by step basis. Typically, all that is known is the outcome of a reaction, a situation that is particularly unenlightening if a synthesis experiment does not produce desirable products. The in situ XRD measurements allow us to observe how the reactants break down, how they might dissolve in a flux, and how new compounds form on both heating and cooling. This information can be used to tailor syntheses in order to avoid unwanted intermediate phases that can compete with the desired compound and to develop alternative synthetic strategies that may be more successful. It can identify underlying mechanisms, such as the redox reaction that is central to the stability of CuO and CuS compounds that will block certain reaction pathways, information that may have more general applicability in related classes of materials. Combining the in situ XRD measurements with calculations of the total energies for both the desired products as well as competing compounds has the potential to take "theory-assisted synthesis" to a new level by connecting realistic synthesis to predictive theory. In the long run, the accumulated knowledge on chemical reactions obtained from in situ XRD measurements could also be used as an input in computation for in silico synthesis to predict viable synthetic routes.

## Conclusions

In this contribution, we have presented our investigation on the $La_2CuO_{4-x}S_x$ quaternary system using combined experimental and computational methods to test if the theoretical compounds $La_2CuO_3S$ and $La_2CuO_2S_2$ could be synthesized. Using in situ XRD measurements, the synthetic obstacle to forming these compounds has been identified, which is the redox reaction between the $Cu^{2+}$- and $S^{2-}$-containing starting materials that drives them away from the desired oxidation states. This incompatibility of the starting materials has also been well described by the DFT calculations, which have shown large, negative reaction energies for these redox reactions. Although the attempts to experimentally realize the theoretical compounds are not successful, this study has shown consistency between experiment and computation. This suggests that one could integrate theory and experiment in a closed loop in exploratory synthesis, where theory could identify theoretical desired materials that are thermodynamically stable, and in situ XRD synthesis could be used to pinpoint the feasible synthetic routes. This new approach may have potential to advance our knowledge on reaction mechanisms involving the formation of extended solids and to accelerate our materials discovery.

**Methods**

1. Synthesis with in situ X-ray diffraction

The starting materials $La_2O_3$, CuO, $La_2S_3$, CuS, $Cu_2O$ and S were purchased from either Alfa Aesar or Aldrich, with stated purity above 99%. $La_2O_2S$ was prepared by mixing stoichiometric $La_2O_3$ and $La_2S_3$ powders, and heating the mixture in a furnace at 900°C overnight. For the in situ high temperature powder X-ray diffraction experiments, stoichiometric powders were mixed and ground thoroughly with a mortar and pestle inside an Argon-filled glove box. The powder mixtures were packed into corundum alumina tubes (OD: 2.39mm, ID: 1.57mm) and two pieces of alumina rods (size matching the ID of the alumina tube) were used to plug the tubes, holding the powder mixture at a fixed position. The setup was then taken out of the glove box and the ends were sealed with alumina-based high-temperature cement (940HT powder and activator from Cotronics Corp). A small amount of air might be introduced into the sample tube before the cement was applied and cured.

The in situ high temperature powder XRD experiments were conducted at beamline 6-ID-D at the Advanced Photon Source (APS), Argonne National Laboratory and at beamline 28-ID-2 at the National Synchrotron Light Source II (NSLS II), Brookhaven National Laboratory. The energy of the synchrotron beam at both facilities was 70 keV, and two types of heating equipment were involved. The furnace for the measurements at beamline 6-ID-D (APS) was constructed by wrapping a hollow alumina core with Kanthal type APM resistance heating wire, providing a 2cm long hot zone. A type R thermocouple was positioned close to the center spot where the beam passed through. The temperature was controlled by a Eurotherm model 2408 PID temperature controller. At beamline 28-ID-2 (NSLS II), the samples were heated by a quadrupole lamp furnace which consisted of four halogen lamps producing a 4mm hot spot at their geometric center. Details of the configurations of this furnace are described elsewhere[35]. In order to establish the temperature at the focal point of the lamps, we inserted a type K thermocouple in the furnace, and aligned the thermocouple and the furnace at the beamline in such a way that the X-ray beam, the thermocouple tip, and the geometric center of the furnace all coincided with each other. A Honeywell UDC3500 universal digital controller was used to control the power output. A linear relationship between the power output and the thermocouple temperature was established, which was then used to determine the sample temperature. However, due to the different absorption of the emitted halogen light radiation between the samples and the thermocouple, the actual sample temperature may deviate from the linear expression. By comparing the temperatures with those of the resistive heating furnace, we found that temperature deviations were no more than 100°C.

The samples were heated up to 950-1100°C at a rate of 20-25°C/min. Diffraction data were collected continuously using a Perkin-Elmer 2D detector in 60s increments, with 1s exposure for each frame, which were then summed over 60s. The samples were heated in place within the resistive furnace, however, those heated in the lamp furnace were slowly oscillated to ensure homogeneous heating, since the length of the samples were ~5mm long, slightly over the size of the hot spot of the lamp furnace. The sample tubes were moved in the following manner: moved to the center of the sample and collected diffraction data for 1 min, and then moved 1mm to the left, collected data for 1 min, and then moved 1mm to right from the center position, collected data for 1min, and repeated the cycle. Since the diffraction patterns appeared to be the same at all three positions, only the data collected at the center position were used for further analysis. The program fit2D was used to integrate the 2D diffraction image[36]. Rietveld refinements were done using the FullProf suite[37]. A sample list and their heating profiles, as well as the details of the refinements are provided in Supplementary Information (Supplementary Table 1 and Fig. 1).

2. Computation methods

We performed density functional theory (DFT) calculations on our target compounds, using the Generalized Gradient Approximation (GGA) as implemented within VASP[38-41] and Projector Augmented Wave (PAW) potentials[42,43]. The relaxation parameters were the defaults from MPRelaxSet in the Materials Project[44] and anion corrections for the oxygen and sulfur containing compounds are those from version 4.5.7 of pymatgen[45]. The remaining total energies for all binary, ternary and quaternary compounds in the La-Cu-O-S chemical system are those computed by the Materials Project[44]. The full list of total energies, listed in the Supplementary Table 3, were used to construct the convex hull[31], which is the surface in the compositional phase space where the total energy is the least, separating stable and unstable compounds. This information is subsequently used to compute reaction energies, and to compare different synthetic routes to those reported in experiments.

**Acknowledgements**

Work at Texas A&M University was supported by the Welch Foundation, grant A-1890-20160319. C.H.Y. and G.K. were supported as part of the Center for Emergent Superconductivity, an Energy Frontier Research Center funded by the US Department of Energy, Office of Science, Office of Basic Energy Sciences under Award No. DEAC0298CH1088. This research used resources of the Advanced Photon Source, a U.S. Department of Energy (DOE) Office of Science User Facility operated for the DOE Office of Science by Argonne National Laboratory under Contract No. DE-AC02-06CH11357. This research used resources of the National Synchrotron Light Source II, a U.S. Department of Energy (DOE) Office of Science User Facility operated for the DOE Office of Science by Brookhaven National Laboratory under Contract No.

DE-SC0012704. H.H. would also like to thank the 6-ID-D beamline staff at the APS and the XPD beamline staff at the NSLS II for their support and useful discussion.

**Author contributions**

H.H., D.E.M, J.W.S., S.Z., M.K., P.K., G.G., A.Z., S.G., J.B., and E.D. performed the in situ XRD experiments, C.H.Y. and G.K. performed the theoretical calculations. H.H. and M.C.A. directed the analysis of the results. H.H., C.H.Y. and M.C.A. contributed to the writing of the paper.

**Competing financial interests**

The authors declare no competing financial interest.

## Figure legends

**Figure 1 Crystal structures of $La_2CuO_4$ and $La_2CuO_2S_2$. a**, Crystal structure of $La_2CuO_4$ ($K_2MgF_4$ type), the parent compound of the first cuprate superconductor. **b**, Crystal structure of the theoretical compound $La_2CuO_2S_2$, which is derived by replacing the apical oxygen in $La_2CuO_4$ with sulfur. Replacing half of the apical oxygen produces $La_2CuO_3S$.

**Figure 2 Synthesis of $La_2CuO_4$ from $La_2O_3$ and CuO powders, monitored by in situ powder X-ray diffraction.** The left panel shows the evolution of the diffraction peaks as temperature changes, and the right panel shows the weight fraction of each phase determined from Rietveld refinements of the diffraction patterns.

**Figure 3 Attempted syntheses for $La_2CuO_3S$, monitored by in situ powder X-ray diffraction. a**, Synthesis from $La_2O_2S$+CuO. **b**, Synthesis from $La_2O_3$+CuS. **c**, Synthesis from 2/3$La_2O_3$+1/3$La_2S_3$+CuO. The left panels show the evolution of the diffraction peaks as temperature changes, and the right panels show the weight fraction of each phase determined from Rietveld refinements of the diffraction patterns. The weight fractions of $La_2O_3$, $La(OH)_3$, and LaOOH are summed and reported as the effective $La_2O_3$ weight fraction in order to make it easier to track the evolutions of phases of interest.

**Figure 4 Non-equilibrium phase diagram of $La_2CuO_{4-x}S_x$ ($0 \leq x \leq 3$).** The phase diagram was obtained from identifying the crystalline products while heating the mixtures of $La_2O_3$, $La_2S_3$, and CuO to high temperatures. The reaction temperatures shown in the figure are contingent on the heating rate ~20°C/min, because the systems were most likely not at equilibria with this relatively fast heating rate. The diffraction data for $x$=4 are not included in the figure since the starting materials involve CuS, which shows different thermodynamic property from CuO. Further details and the reaction pathways are provided in Supplementary Table 2.

**Figure 5 Thermodynamic instability of the theoretical compounds $La_2CuO_2S_2$ and $La_2CuO_3S$. a**, Convex hull of the La-Cu-O-S quaternary system. **b**, Energy above hull of $La_2CuO_2S_2$ and $La_2CuO_3S$ in comparison with those occurring in syntheses. The grey region, ±50meV/atom, shows the probable inherent uncertainty from computation.

Table 1 Comparison of reaction energies of the proposed and the experimental reaction pathways.

| Reaction | Computed energy (meV/atom) |
|---|---|
| Parent compound: $La_2O_3$ + CuO = $La_2CuO_4$ | 51[ref.17] |
| Proposal A | |
| $La_2O_2S$ + CuO = $La_2CuO_3S$ | 182 |
| Experiment A | |
| Step 1 (760-1000 °C): $La_2O_2S$ + 8 CuO = 4 $Cu_2O$ + $La_2O_2SO_4$ | –197 |
| Step 2 (1000-1100 °C): 2 $La_2O_2S$ + $Cu_2O$ = 2 LaCuSO + $La_2O_3$ | –54 |
| Total: $La_2O_2S$ + CuO = 7/8 LaCuSO + 7/16 $La_2O_3$ + 1/8 $La_2O_2SO_4$ + 1/16 $Cu_2O$ | –118 |
| Proposal B | |
| $La_2O_3$ + CuS = $La_2CuO_3S$ | 266 |
| Experiment B | |
| Step 1 (450-760 °C): 8 CuS + 4 $La_2O_3$ = 4 $Cu_2S$ + 3 $La_2O_2S$ + $La_2O_2SO_4$ | –43 |
| Step 2 (860-1000 °C): $La_2O_2S$ + $Cu_2S$ = 2 LaCuSO | –22 |
| Total: $La_2O_3$ + CuS = 1/8 $La_2O_2SO_4$ + 1/8 $Cu_2S$ + 3/4 LaCuSO + 1/2 $La_2O_3$ | –37 |
| Proposal C | |
| 2/3 $La_2O_3$ + 1/3 $La_2S_3$ + CuO = $La_2CuO_3S$ | 79 |
| Experiment C | |
| Step 1 (600-760 °C): 2 $La_2O_3$ + $La_2S_3$ + 24 CuO = 3 $La_2O_2SO_4$ + 12 $Cu_2O$ | –231 |
| or $La_2S_3$ + 2 $La_2O_3$ = 3 $La_2O_2S$ | –144 |
| $La_2O_2S$ + 8 CuO = 4 $Cu_2O$ + $La_2O_2SO_4$ | –197 |
| Step 2 (760-950 °C): 2 $La_2O_2S$ + $Cu_2O$ = 2 LaCuSO + $La_2O_3$ | –54 |
| Total: 2/3 $La_2O_3$ + 1/3 $La_2S_3$ + CuO = 1/8 $La_2O_2SO_4$ + 7/8 LaCuSO + 7/16 $La_2O_3$ + 1/16 $Cu_2O$ | –221 |

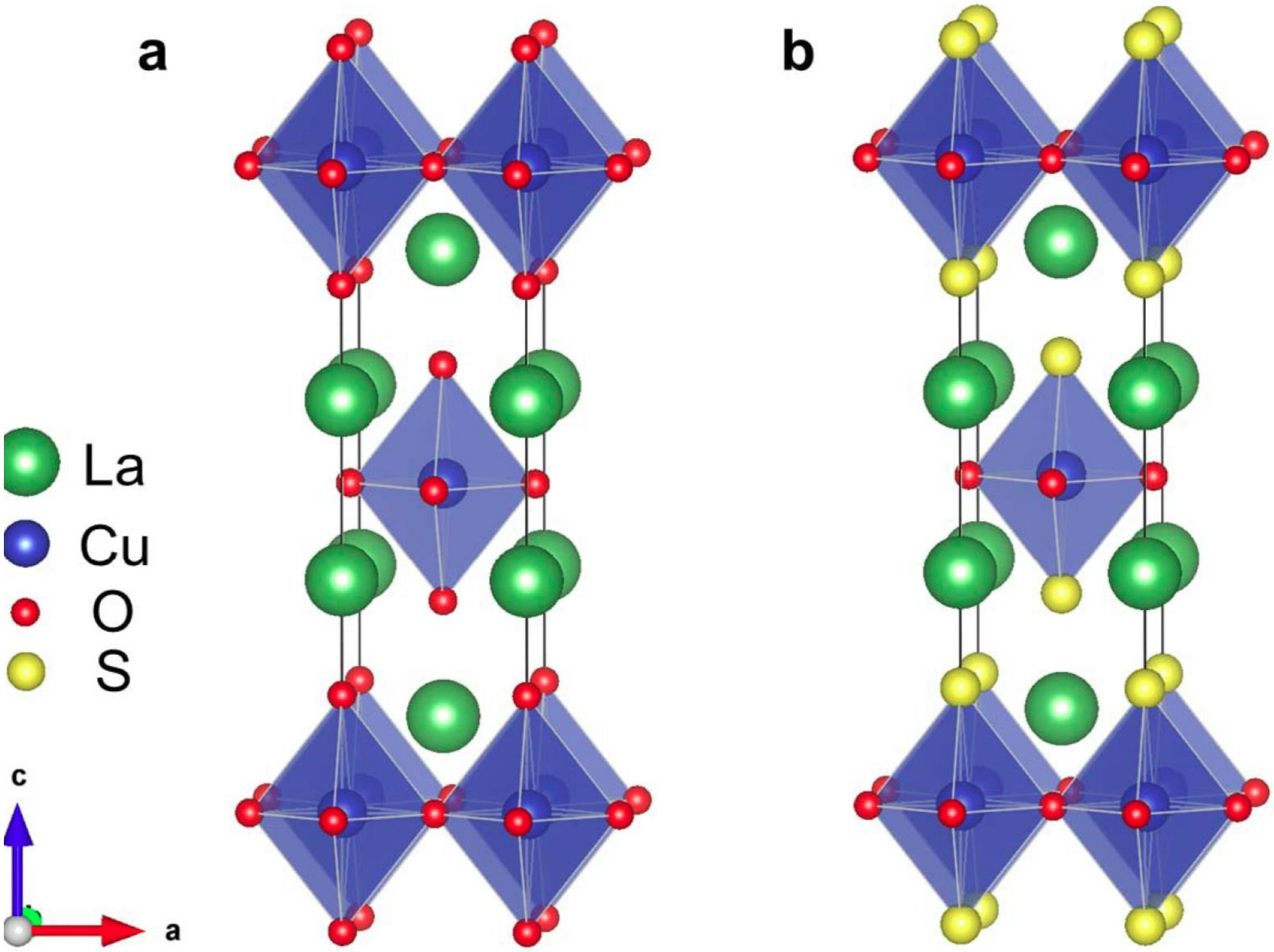

Figure 1.

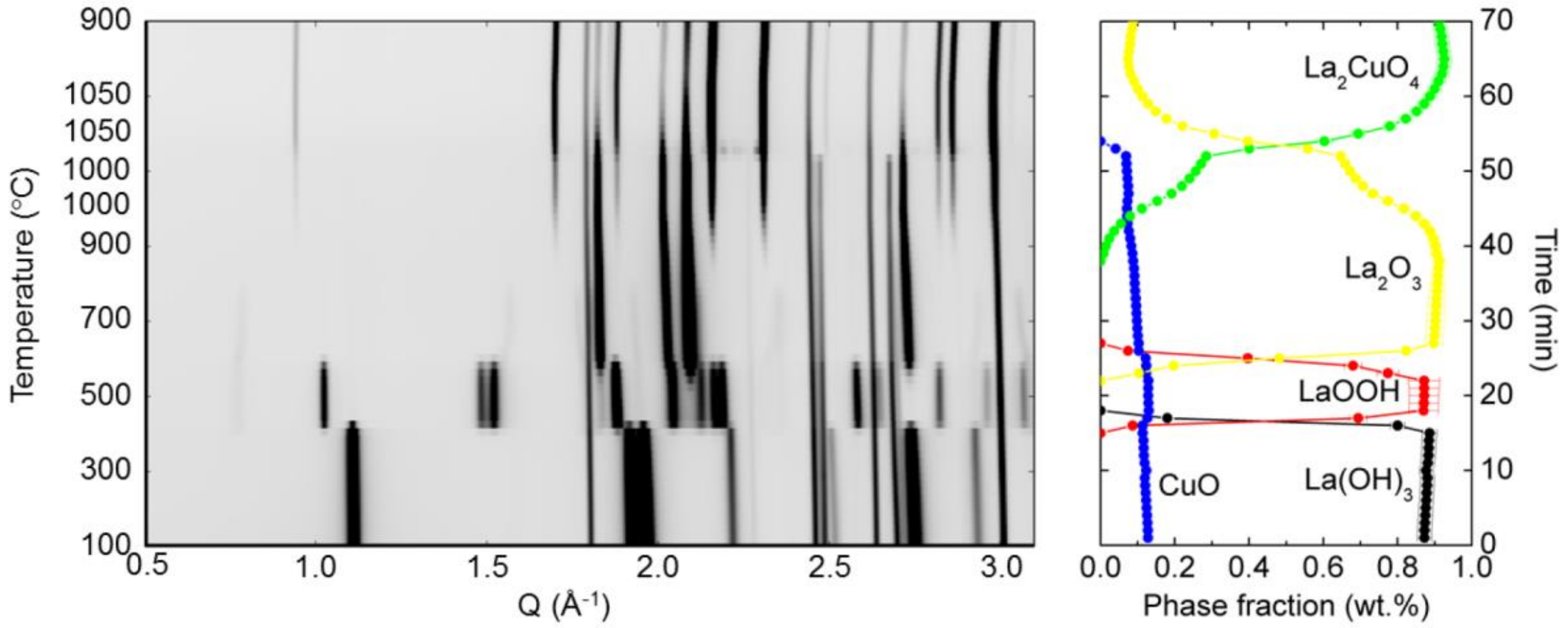

Figure 2

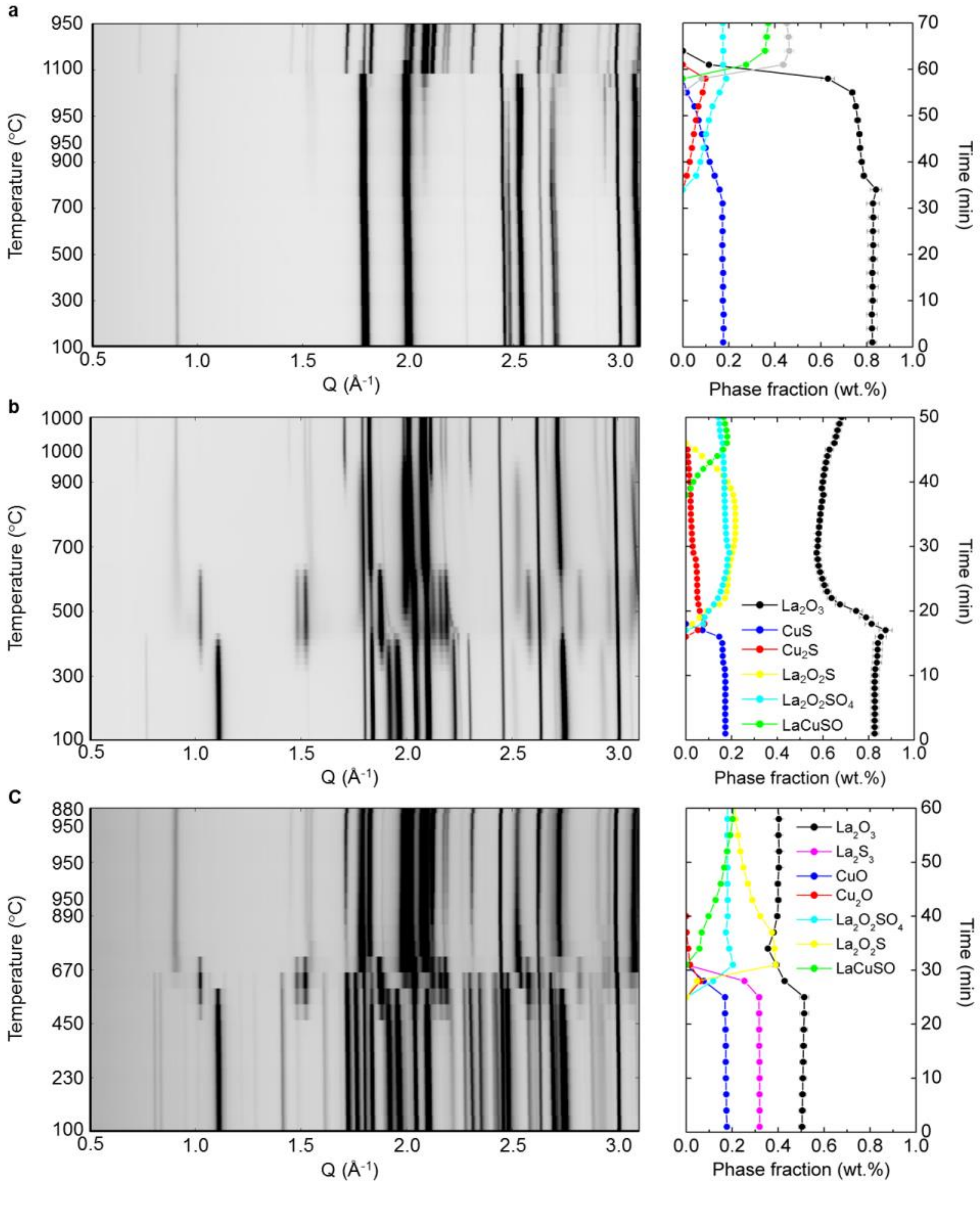

Figure 3

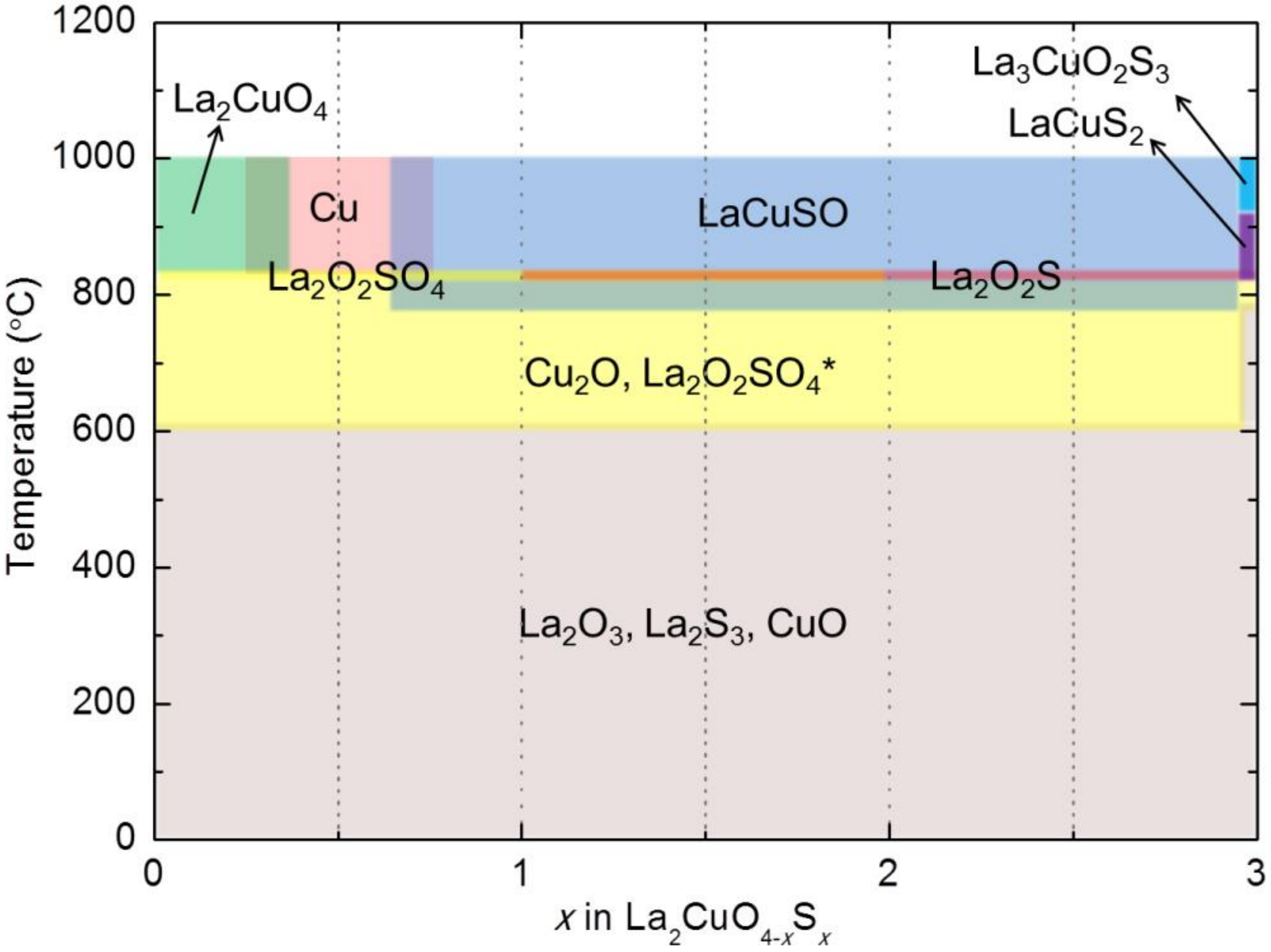

Figure 4

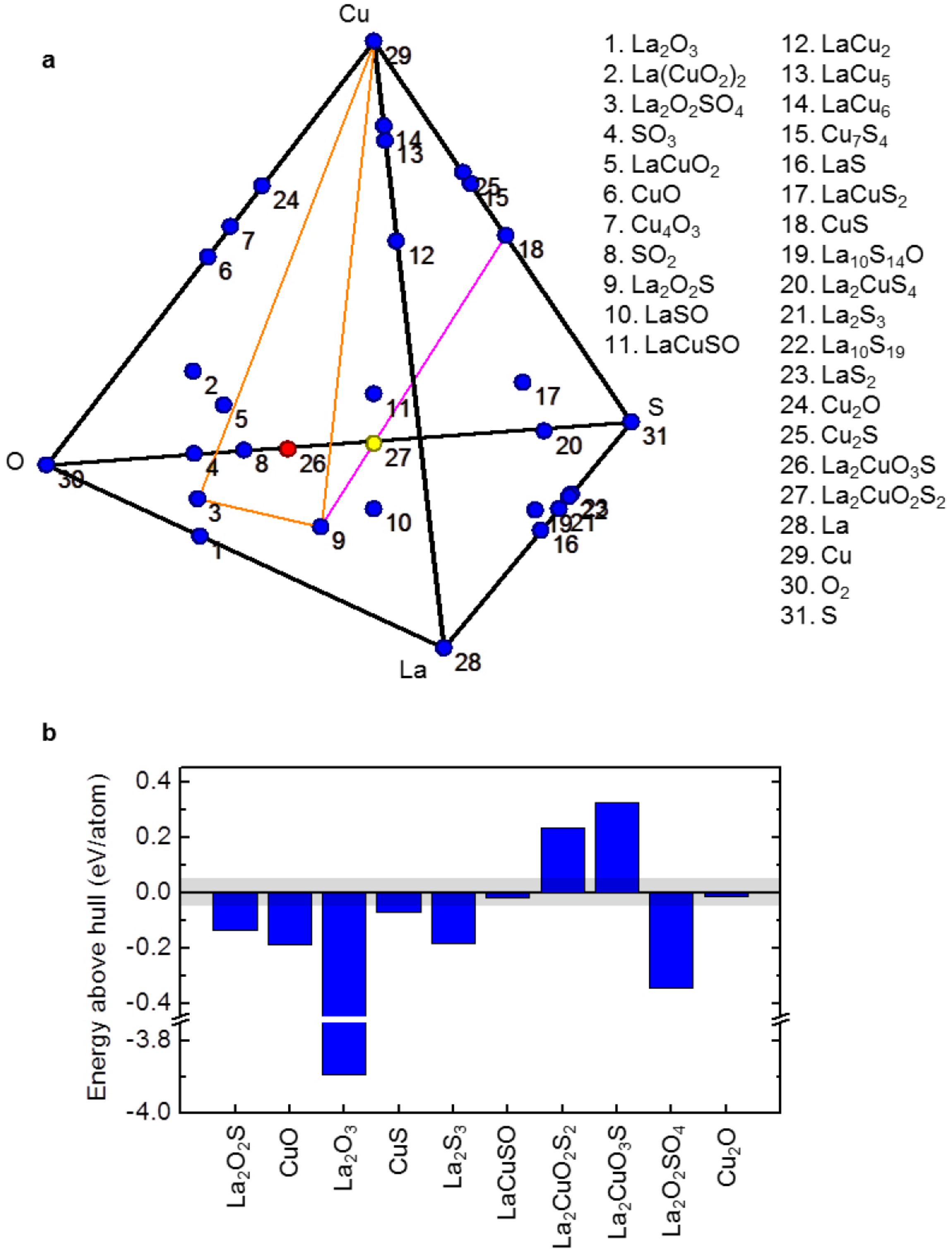

Figure 5